\documentclass[pra,preprint,showpacs]{revtex4}
\usepackage{graphicx}
\usepackage{amssymb}
\usepackage{subfigure}
\usepackage{color}
\newcommand{\cor}{\textcolor{blue}}

\begin{document}

\title{Ultrashort light bullets described by the two-dimensional sine-Gordon equation}
\author{Herv{\'e} Leblond$^{1}$  and  Dumitru Mihalache$^{2,3}$ }
\affiliation{
$^{1}$Laboratoire de Photonique d'Angers, EA 4464
 Universit\'e d'Angers, 2 Bd. Lavoisier, 49045 Angers Cedex 01, France\\
$^{2}$Horia Hulubei National Institute for Physics and Nuclear Engineering (IFIN-HH),
407 Atomistilor, Magurele-Bucharest, 077125, Romania\\
$^{3}$Academy of Romanian Scientists, 54 Splaiul Independentei, Bucharest 050094, Romania}

\begin{abstract}

By using a reductive perturbation technique applied to a two-level model, a 
generic two-dimensional sine-Gordon evolution equation governing the propagation of femtosecond spatiotemporal optical solitons in Kerr  media beyond the slowly-varying envelope approximation is put forward.
Direct numerical simulations show 
that, in contrast to the long-wave approximation, no collapse occurs, and that
robust (2+1)-dimensional ultrashort light bullets may form from adequately chosen few-cycle input spatiotemporal waveforms. In contrast to the case of quadratic nonlinearity, the light bullets oscillate in both space and time,
and are therefore not steady-state lumps.

\end{abstract}

\pacs{42.65.Tg, 42.65.Re, 05.45.Yv}

\maketitle

\section{Introduction}

Since the first experimental realization more than one decade ago of two-cycle optical pulses 
in mode-locked Ti-sapphire lasers using
double-chirped mirrors \cite{two-cycle1}-\cite{two-cycle3} the field of few-cycle pulses (FCPs) has grown
into one of the major area of modern ultrafast optics. These studies led to remarkable achievements in the field of extreme nonlinear optics, time-resolved laser spectroscopy, generation of soft X-ray radiation and isolated and controlled attosecond light pulses in the extreme ultraviolet regime, thereby opening the door for
real-time probing of fast electron dynamics associated with excited-state atomic and molecular processes 
\cite{review}-\cite{atto2}. These FCPs are key elements to many time-domain applications, driving the quest for stable mode-locked laser sources with ultrahigh spectral bandwitdth. A remarkable recent advance in this area was the syntesis of a single cycle of light in the near-infrared regime with compact erbium-doped fiber
technology \cite{Krauss_Nat_Phot_2010}. Another recent achievement is the possibility to use different optical parameter amplifiers to produce FCPs over a broad frequency range. 
The optical parametric chirped pulse amplification method has become one of the
standard procedures to generate high peak power pulses with durations of a few optical cycles.
Two- and three-cycle pulses with carrier wavelengths almost continuously tunable from visible to mid-infrared were recently generated \cite{OPA,OPA_OE}. 

On the theoretical arena the studies of the unique physics of FCPs concentrated on three main directions: (i) the quantum approach \cite{tan08,ros07a,ros08a,naz06a}, (ii) the refinements within the framework of the slowly varying envelope approximation (SVEA) of the nonlinear Schr\"{o}dinger-type envelope equations \cite{Brabek_PRL,Tognetti,Voronin,Kumar}, and non-SVEA models \cite{quasiad,leb03,igor_jstqe,igor_hl,interaction}.  
In Kerr-like media the physics of (1+1)-dimensional FCPs  can be adequately described beyond the SVEA by using different dynamical models, such as the modified Korteweg-de Vries (mKdV) \cite{quasiad}, sine-Gordon (sG) \cite{leb03,igor_jstqe}, or mKdV-sG equations \cite{igor_hl,interaction}.
The above mentioned evolution equations admit breather solutions, which are most suitable for describing the physics of few-optical-cycle solitons. 
Notice that other non-SVEA models \cite{kozlov97,bespalov,berkovsky}, especially the so-called short-pulse equation \cite{sch04a}, have been proposed.
Moreover, a non-integrable generalized Kadomtsev-Petviashvili (KP) equation \cite{KP} (a two-dimensional version of the mKdV model) was also put forward for describing the (2+1)-dimensional few-optical-cycle spatiotemporal soliton propagation in cubic nonlinear media beyond the SVEA~\cite{igor1,igor_matcom}.
However, it has recently been shown that collapse of FCPs occurs for high enough intensities in the frame of cubic generalized Kadomtsev-Petviashvili model, which describes ultrashort spatiotemporal optical pulse propagation in cubic (Kerr-like) media \cite{fcp_collapse}.
Notice that of particular interest for the physics of the (1+1)-dimensional FCPs is the mKdV-sG equation; thus by using a system of two-level atoms, it has been shown in Ref. \cite{igor_hl} that the propagation of ultrashort pulses in Kerr optical media
is fairly well described by a generic mKdV-sG equation, which was also derived and studied in Refs. ~\cite{saz01b,bugay}. Recently, we showed that the mKdV-sG model is the most general of all approximate non-SVEA
models for FCPs,  and in fact contains all of them \cite{Leblond_Mihalache_PRA_2009}.
Also, the propagation of single-cycle gap solitons in subwavelength periodic structures, without the use of the SVEA of the Maxwell-Bloch equations has also been theoretically predicted in a recent study~\cite{xie10}.

As concerning the physical assumptions made in order to get the above mentioned generic nonlinear evolution equations we first mention that soliton propagation implies that damping can be neglected. In dielectric media, this occurs far from 
any resonance frequency. Let us consider a two-level model with characteristic frequency $\Omega$, and denote by 
$\omega_w$  a frequency characteristic for the FCP soliton under consideration. The transparency condition implies that 
either $\omega_w\ll\Omega$  or $\Omega\ll\omega_w$. 
The former case ($\omega_w\ll\Omega$) corresponds to the {\it long wave approximation}, whereas the latter case corresponds to the {\it short wave approximation}.  
In the frame of a two-level model, a mKdV equation is obtained if 
the frequency of the transition $\Omega$ is far above the characteristic wave frequency $\omega_w$ (the long-wave approximation regime),
while a SG model is valid if $\Omega$ is much smaller than $\omega_w$ (the short-wave approximation regime).
However, some of the transitions 
frequencies $\Omega_j$ are much smaller than $\omega_w$, and the other ones much larger than $\omega_w$. 
This more realistic situation can be modeled by considering two transitions only, with 
different frequencies $\Omega_1$ and $\Omega_2$. The physical system is thus equivalent 
to a two-component medium, each of the two components being described by a two-level model. As a result, a mKdV-sG model was put forward \cite{igor_hl,interaction,saz01b,bugay}, which is completely integrable in certain particular cases by means of the inverse scattering transform. It admits stable solutions of 'breather' type, which also give a good description of few-optical-cycle soliton propagation.

The propagation of FCPs in a quadratic medium has also been described by either a KdV or a KP equation, in (1+1) or (2+1) dimensions, respectively,  which evidenced either the stability of a plane wavefront, for a normal dispersion, or the formation of a localized spatiotemporal half-cycle soliton, for an anomalous dispersion \cite{kdvopt,LKM_KP}.
By using a multiscale analysis, a 
generic KP evolution equation governing the propagation of femtosecond spatiotemporal optical solitons in quadratic nonlinear media beyond the SVEA was recently put forward \cite{LKM_KP}. 
Direct numerical simulations showed the formation, from adequately chosen few-cycle input pulses, of both stable line solitons (in the case of a quadratic medium with normal dispersion) and of stable lumps (for a quadratic medium with anomalous dispersion). 
The perturbed unstable line solitons decay into stable lumps for a quadratic nonlinear medium with anomalous dispersion \cite{LKM_KP}. However, in Ref. \cite{LKM_KP} we considered a set of two-level atoms and we have assumed that
the characteristic frequency $\omega_w$ of the considered electromagnetic wave in the optical spectral range is much less than the transition frequency $\Omega$ of the atoms, i.e., we worked in the so-called long-wave approximation regime.

Summarizing, in the {\it long wave approximation}, for a quadratic nonlinearity, half-cycle light bullets in the form of a single lump may exist, while for a cubic nonlinearity collapse occurs. In the present work we will show that in the {\it short wave approximation} for a cubic nonlinearity, a third physical situation occurs: few-cycle light bullets may form, oscillating in both space and time.

More precisely, the aim of this paper is to derive a generic partial differential equation describing the dynamics of (2+1)-dimensional spatiotemporal solitons beyond the SVEA model equations in the so-called short-wave approximation regime, when we consider the situation in which the resonance frequency $\Omega$ of the two-level atoms is below the characteristic optical frequency $\omega_w$. 
\cor{Several generalizations to (2+1) dimensions of the sG equation exist in the literature and can be derived in the short-wave approximation; some of them support localized solitons which are not oscillating as the sG breather and one-dimensional FCP solitons in Kerr media~\cite{ferro_prl,ferro_prb,swfer_let,swfer_perp}. The question whether the (2+1)-dimensional generalization of sG equation, which is valid for FCP propagation in cubic nonlinear media, is one of the nonlinear dynamical systems discussed in Refs. \cite{ferro_prl,ferro_prb,swfer_let,swfer_perp} or some other one, can only be decided by means of a rigorous derivation starting from the basic equations.}
We use a reductive perturbation method (multiscale analysis) \cite{tutorial} and as a result we get a generic two-dimensional sine-Gordon (2D sG) equation to describe the propagation of ultrashort (2+1)-dimensional light bullets in a system of two-level atoms beyond the SVEA.  
\cor{This is the first result of the present work.}
 
This work is organized as follows. In Sec. II we derive the 2D sG equation governing the propagation of femtosecond spatiotemporal solitons beyond the SVEA. Direct numerical simulations of the 2D sG nonlinear partial differential equation are given in Sec. III, where it is shown the generation of stable (2+1)-dimensional FCPs from perturbed FCP plane waves.
\cor{This is the second result of this paper. 
We note that a numerical resolution of the 2D sG equation was published in an earlier work \cite{xin00}, which showed the existence of ultrashort light bullets. However, it was believed that their amplitudes were not constant but decreased during propagation. We prove numerically that this decay is only a transient feature corresponding to a reshaping of the input, and that a stable steady-state do exist. In other words we prove that the localized pulse of Ref. \cite{xin00} is really an ultrashort light bullet. This is the third result of our paper.}

 Section IV presents our conclusions.

\section{Derivation of the two-dimensional sine-Gordon equation}

We consider a set of two-level atoms with the Hamiltonian
\begin{equation}
H_0=\hbar\left(\begin{array}{cc}\omega_a&0\\0&\omega_b\end{array}\right),\label{hamil}
\end{equation}
where $\Omega=\omega_b-\omega_a>0$ is the frequency of the transition.
The evolution of the electric field $E$ 
 is described by the wave equation
\begin{equation}
\left(\partial_y^2+\partial_z^2\right)E=\frac1{c^2}\partial^2_t\left(E+4\pi P\right),\label{max}
\end{equation}
where $P$ is the polarization density.
 The light propagation is
 coupled with  the medium by means of a dipolar electric momentum \begin{equation}
\mu=\left(\begin{array}{cc}0&\mu\\\mu^\ast&0\end{array}\right)
\end{equation} directed along the same direction
 $x$ as the electric field, according to
 \begin{equation}
 H=H_0-\mu E,\label{eq17}
 \end{equation}
 and the polarization density $P$ along the $x$-direction is
 \begin{equation}
 P=N \mathrm{Tr}\left(\rho\mu\right),\label{eq18}
 \end{equation}
 where $N$ is the volume density of atoms and $\rho$ the density matrix.
Since, as shown in Ref. \cite{leb03}, the relaxation terms containing the relaxation times for both the level 
populations and coherences
can be neglected, due to the fact that the relaxation occurs very slowly with regard to optical oscilations.
Thus the density-matrix evolution equation (Schr{\"o}dinger equation) reduces to
 \begin{equation}
 i\hbar\partial_t\rho=\left[H,\rho\right].\label{schr}\label{eq19}
 \end{equation}

Transparency implies that the characteristic frequency $\omega_w$ of the considered
radiation (in the optical range)  strongly differs from  the resonance frequency $\Omega$ of the atoms.
Here we consider the so-called short-wave approximation, by assuming that $\omega_w$ is much larger than $\Omega$.

Hence $\Omega$ is small, which can be expressed as 
\begin{equation}
\Omega=\varepsilon \tilde \Omega,\label{omeg}
\end{equation}
$\varepsilon$ being a small parameter.
It is convenient to rescale the time unit in the basic 
equations according to (\ref{omeg}), and hence we set
\begin{equation}
c=\varepsilon \tilde  c,\quad H=\varepsilon \tilde  H, \quad t=\tilde t/\varepsilon,
\end{equation}
and analogously for $\omega_a$, $\omega_b$, $H_0$.
The set of equations (\ref{hamil})-(\ref{eq19}) is invariant under this transform, except that Eq. (\ref{eq17}) becomes
 \begin{equation}
 \tilde  H=\tilde  H_0-\frac\mu\varepsilon E.\label{eq17b}
 \end{equation}
It is equivalent to state that $\Omega $ is small in the original units or that the typical wave frequency
$\tilde\omega_w$, with respect to the rescaled time $\tilde t$, becomes very large, and corresponds
thus to a fast variable
\begin{equation}
\tau=\frac1\varepsilon\left(\tilde t-\frac z{\tilde V}\right).\label{scal1}
\end{equation}
Notice that we could equivalently have written $\tau=t-z/V$, as a zero order variable in the original units. However, the physical interpretation of
the approximation and its link to the standard short-wave approximation as defined in \cite{man1,man2,tutorial}
would be much less clear. 
The delayed time $\tau$ involves propagation at some speed $\tilde V$ to be determined.
It is assumed to vary rapidly in time according to the assumption $\omega_w\gg \Omega$.
This motivates the introduction of the variables
\begin{equation}
\zeta=z,\quad
\eta=\frac y{\sqrt{\varepsilon}}.\label{scal}
\end{equation}
The pulse shape described by the variable $\tau$ evolves more slowly in time, the
corresponding scale being that of variable $\zeta$.
The transverse spatial variable $y$ has an intermediate scale, comparable to what is usually considered
 in long-wave approximations \cite{tutorial}.

Next we use the reductive perturbation method as developed in Ref. \cite{tutorial}. To this aim we expand the electric field $E$ as power series of a small parameter $\varepsilon$:
\begin{equation}
E=E_0+\varepsilon E_1+\varepsilon^2E_2+\varepsilon^3E_3+\ldots.\label{exp}
\end{equation}
The polarization density $P$ and the density matrix $\rho$ are expanded in the same way.
The expansion [Eqs. (\ref{scal}) and (\ref{exp})] is then reported into the basic 
 equations (\ref{hamil})-(\ref{eq19}), and solved order by order.
The computation follows the same steps as the (1+1)-dimensional case, see Ref. \cite{leb03}.

At order $1/\varepsilon$, the Schr\"odinger equation (\ref{schr}) reduces to
\begin{equation}
i\hbar\partial_\tau\rho_0=-E_0\left[\mu,\rho_0\right].\label{schr-1}
\end{equation}
We label the components of any Hermitian matrix $u$ as
\begin{equation}u=\left(\begin{array}{cc}u_a&u_t\\u_t^\ast&u_b\end{array}\right).
\end{equation}
We assume that the coherences are zero long before the pulse, i.e., as $z\longrightarrow +\infty$ or $\tilde t\longrightarrow -\infty$
\begin{equation}
\lim _{\tau \longrightarrow -\infty}\rho_{j,t}=0
\end{equation}
for any $j\geq 0$.
Then the off-diagonal terms of Eq. (\ref{schr-1}) allow to express the coherence $\rho_{0,t} $ as 
\begin{equation}
\rho_{0,t}=\frac{i\mu}{\hbar}\int_{-\infty}^\tau E_0w_0 d\tau',
\end{equation}
where $w_0=\rho_{0,b}-\rho_{0,a}$ is the population inversion.
We do not assume a pumping, hence 
\begin{equation}
\lim _{\tau \longrightarrow -\infty}w_0=w_{th},\label{clt}
\end{equation}
in which $-1<w_{th}<0$ is the population inversion at thermodynamic equilibrium. If all atoms are initially in the fundamental state, then 
$w_{th}=-1$. The higher order terms of the population inversion
 ($w_j=\rho_{j,b}-\rho_{j,a}$ with $j\geq 1$) vanish at infinity.
Using the above boundary conditions and definitions, we get from the diagonal terms of (\ref{schr-1})
the evolution equation for $w_0$, as
\begin{equation}
\partial_\tau w_0=\frac{-4\left|\mu\right|^2}{\hbar^2}E_0\int_{-\infty}^\tau E_0w_0 d\tau'. \label{sg2d}
\end{equation}
Since the polarization density at this order is
$P_0=N\left(\rho_{0,t}\mu^\ast+c.c.\right)=0$, where c.c. denotes the complex conjugate, we see from the wave equation (\ref{max}) at
leading order $1/\varepsilon^2$ that the velocity $\tilde V$ should be $\tilde V=\tilde  c$ up to this order.
Eq. (\ref{schr}) at order $\varepsilon^0$ is
\begin{equation}
i\hbar\partial_\tau\rho_1=\left[\tilde  H_0,\rho_0\right]-E_0\left[\mu,\rho_1\right]-E_1\left[\mu,\rho_0\right].\label{schr0}
\end{equation}
The off-diagonal terms yield
\begin{equation}
\rho_{1,t}=i\tilde \Omega\int_{-\infty}^\tau\rho_{0,t}d\tau'+\frac{i\mu}{\hbar}\int_{-\infty}^\tau\left( E_0w_1+E_1w_0\right) d\tau'.
\end{equation}
It allows to compute the leading term of the polarization density 
$P_1=N\left(\rho_{1,t}\mu^\ast+c.c.\right)$, as
\begin{equation}
P_1=\frac{-2N\tilde \Omega\left|\mu\right|^2}{\hbar}\int_{-\infty}^\tau\int_{-\infty}^{\tau'}E_0w_0 d\tau''d\tau'.
\end{equation}
Reporting $P_1$ into the wave equation (\ref{max}) at order $1/\varepsilon$ gives the evolution equation
for the electric field
\begin{equation}
\left(\partial_\eta^2-\frac2{\tilde  c}\partial_\zeta\partial_\tau\right)E_0=
\frac{-8\pi N\tilde \Omega\left|\mu\right|^2}{\hbar \tilde  c^2}E_0w_0,\label{sg1d}
\end{equation}
which together with Eq. (\ref{sg2d}) yields the sought nonlinear system of equations for $E_0$ and $w_0$.

Eqs. (\ref{sg2d}) and (\ref{sg1d}) can be put in the normalized form
\begin{eqnarray}
B_{ZT}&=&AB+B_{YY},\label{sys2d1}\\
A_T&=&-BC\label{sys2d2},\\
C_T&=&AB\label{sys2d3},
\end{eqnarray}
by setting 
\begin{equation}
Y=\frac y{l_r},\quad Z=\frac z{L_r},\quad T=\frac1{T_r}\left(t-\frac zc\right),\quad A=\frac {w_0}{w_r},\quad B=\frac{E_0}{E_r},
\end{equation}
with $L_r$ a reference length in the micrometer range, $l_r=L_r/\sqrt2$, $T_r=L_r/c$,
\begin{equation}
w_r=\frac{\hbar c^2}{4\pi N\Omega|\mu|^2L_r^2},
\end{equation}
and
\begin{equation}
E_r=\frac{\hbar c}{2|\mu|L_r}.
\end{equation}
The boundary conditions, taking into account Eq. (\ref{clt}) 
are then
\begin{equation}\lim_{T\longrightarrow -\infty}C=0,\quad \lim_{T\longrightarrow -\infty}A=\frac{w_{th}}{w_r}.\label{bc_ac}
\end{equation}

The coupled system of partial differential equations (\ref{sys2d1})-(\ref{sys2d3}) is a two-dimensional generalization 
of the sine-Gordon (sG) equation. It is worth to compare it 
 to the set of equations derived by the same reductive perturbation method in the case of electromagnetic wave propagation in ferromagnetic media
 \cite{ferro_prl,ferro_prb,swfer_let,swfer_perp}; however a careful analysis shows that system (\ref{sys2d1})-(\ref{sys2d3}) cannot be reduced to any of these equations.

Next it is seen from Eqs. (\ref{sys2d2})-(\ref{sys2d3}) that $\partial_T\left(A^2+C^2\right)=0$. This allows us to introduce a function
$\psi=\psi(Z,T)$ as
\begin {equation}
 A=U\cos\psi,\quad C=U\sin\psi,\label{psin}
\end{equation}
where $U$ depends on $Z$ only. From the boundary conditions (\ref{bc_ac}), we see that 
\begin{equation}\lim_{T\longrightarrow -\infty}\psi=0\label{bc_psi}\end{equation}
and thus $U=w_{th}/w_r$, and is independent of $Z$ (except if inhomogeneous pumping is present, but this situation is excluded here, see Eq. (ref{clt}).
Reporting Eq. (\ref{psin}) into (\ref{sys2d2}) and (\ref{sys2d3}) shows that $\psi_T=B$. 
Then, after integration using (\ref{bc_psi}),  Eq. (\ref{sys2d1}) reduces to 
\begin{equation}
\psi_{ZT}=U \sin\psi +\psi_{YY},\label{sineg2d}
\end{equation}
which is known as the two-dimensional sG (2D-sG) equation.
\cor{Notice that, to the best of our knowledge, the 2D sG equation was never derived before  
by means of the short-wave formalism in any physical context}.
\section{Numerical resolution of 2D sG}

\begin{figure}
\begin{center}
\includegraphics[width=7cm]{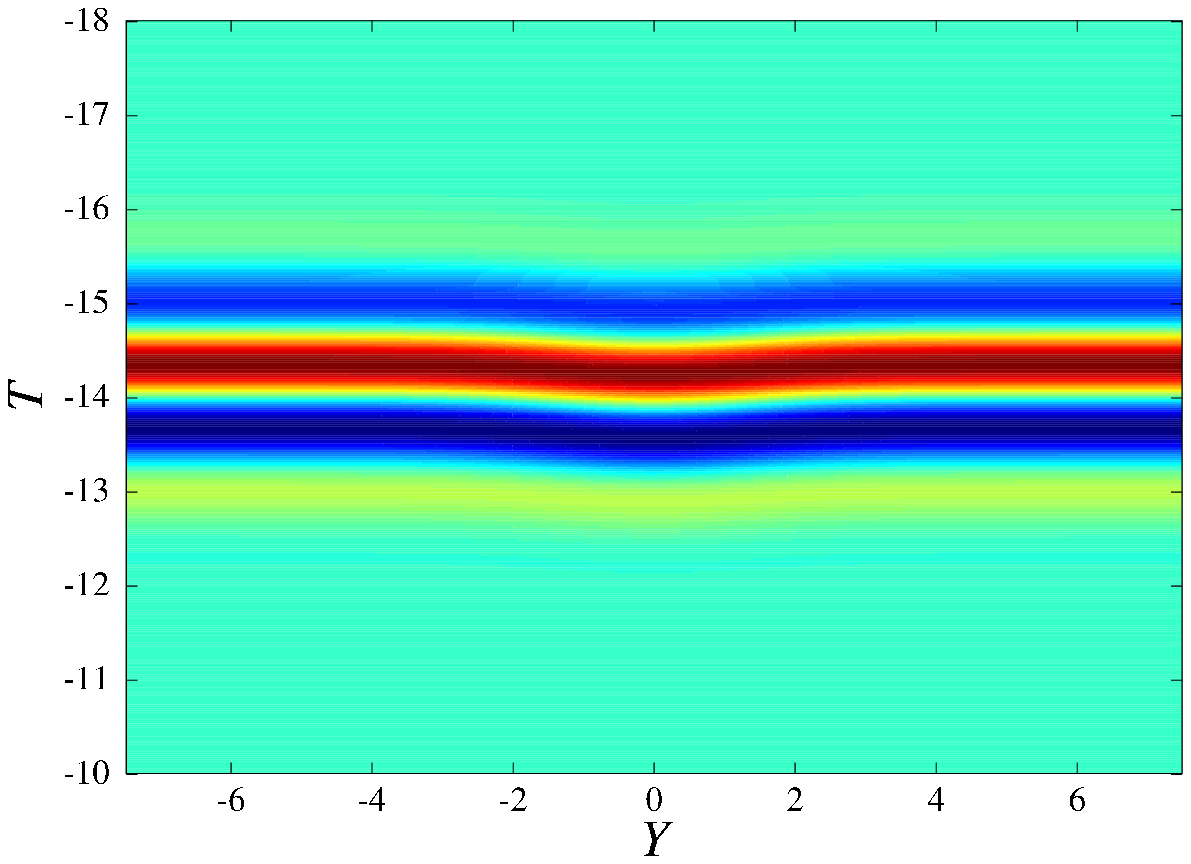}
\includegraphics[width=7cm]{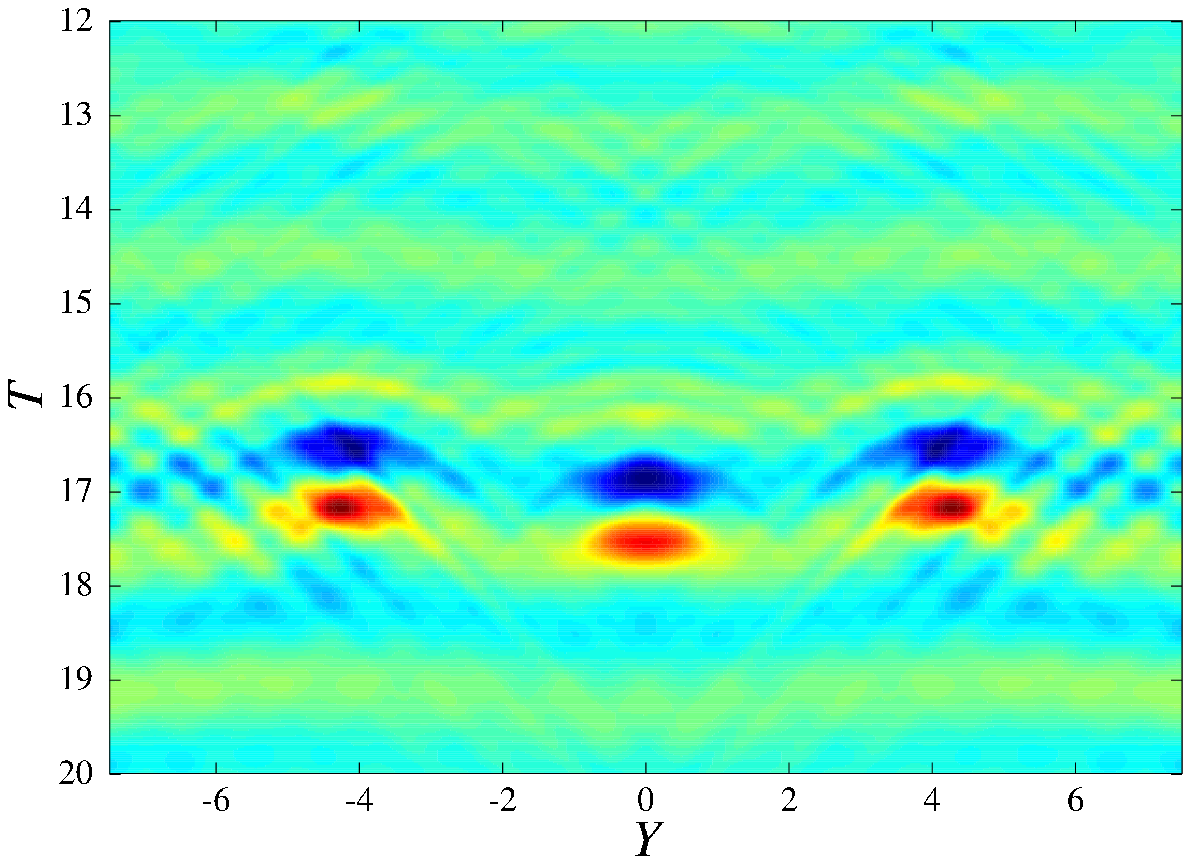}
\caption{ (Color online) Evolution of a perturbed input FCP plane wave into two dimensional FCP solitons. a) Input ($Z=0$), 
 b) $Z= 52.2$.}
\label{sg_if}
\end{center}\end{figure}

\begin{figure}
\begin{center}
\includegraphics[width=7cm]{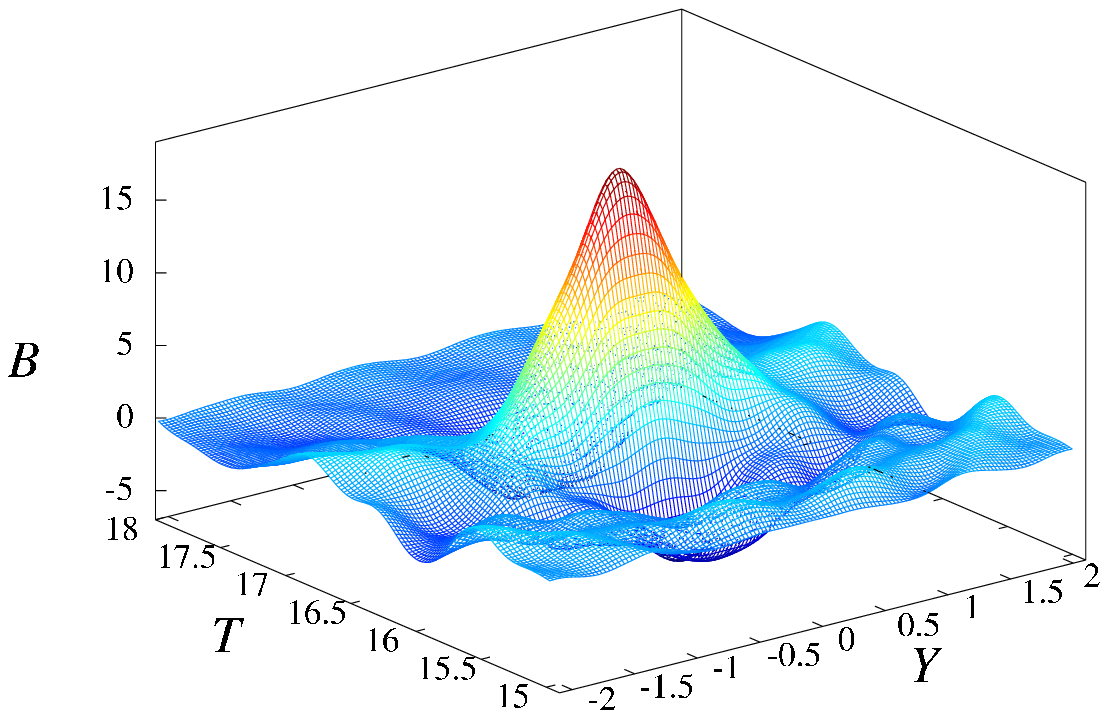}
\includegraphics[width=7cm]{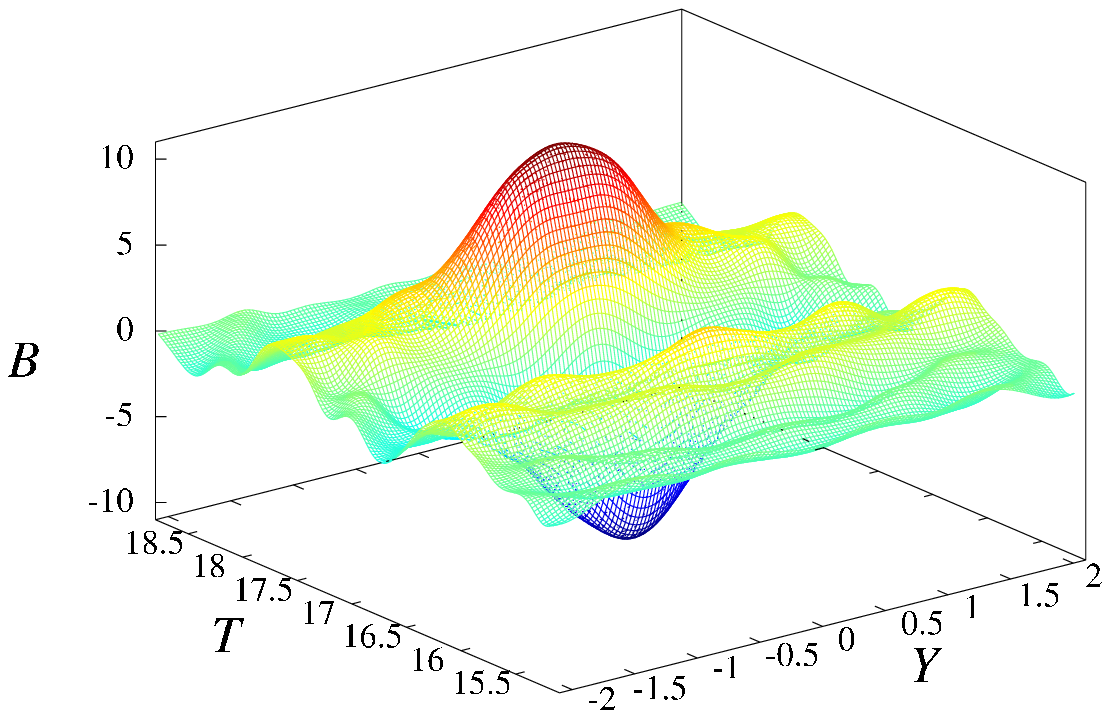}
\caption{(color online) Two stages of the oscillations of the two dimensional FCP formed in the computation shown in Fig. \ref{sg_if}.
a) $Z=51$, 
 b) $Z= 52.2$. 
}
\label{sg_fin_mesh}
\end{center}\end{figure}

The 2D sG equation is solved in the form (\ref{sys2d1})-(\ref{sys2d3}), using a simplified version of the  numerical scheme given in Refs. \cite{ferro_prl,ferro_prb,swfer_let,swfer_perp}.
Starting from an initial data in the form of a FCP plane wave with temporal shape as
\begin{equation}
 B=\beta \exp\left[-\frac{(T-T_1-T_0)^2}{w_T^2}\right]\cos\left[\frac{2\pi}{\theta}(T-T_1)+\pi/2\right],
\end{equation}
in which \begin{equation}
T_1=0.1\exp\left(-\frac{Y^2}{w_Y^2}\right)
\end{equation}
yields a transverse perturbation.
The parameters effectively used in computation were			
$w_T= 1$, $ w_Y=2$, $T_0 =-0.2$, 
$\theta= 1.5$, $\beta= 8$, and $U=-10$. 
The numerical resolution clearly shows that, during propagation, we get the formation of localized solitons (see Fig. \ref{sg_if}).
The two-dimensional FCP solitons are oscillating structures, which are localized in both space and time, see Fig. \ref{sg_fin_mesh}.

\begin{figure}
\begin{center}
\includegraphics[width=7cm]{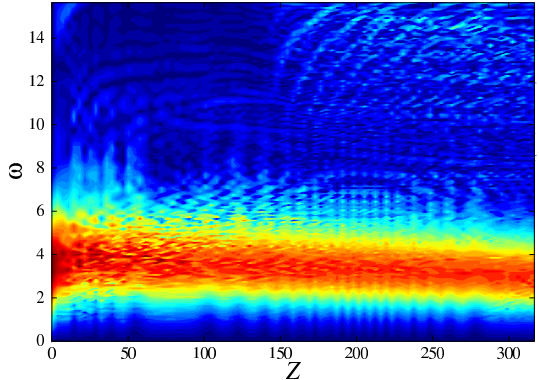}
\caption{(Color online) Evolution of the optical spectrum of the center of the pulse.}
\label{fig_spec}
\end{center}\end{figure}

\begin{figure}
\begin{center}
\includegraphics[width=7cm]{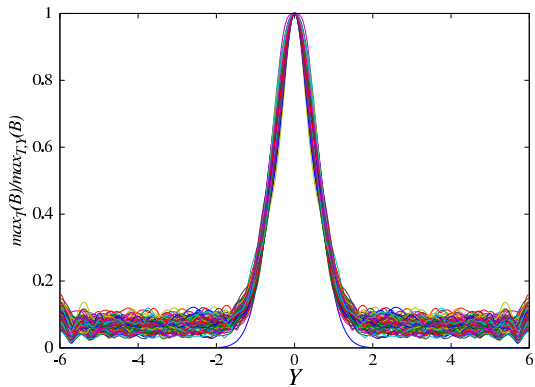}
\caption{(Color online) Evolution of the pulse width. The quantity $\max_T (B)/\max_{T,Y}(B)$ is plotted versus $Y$ 
for values of $Z$ form  0 to 317.}
\label{fig_larg}
\end{center}\end{figure}

The  two-dimensional FCP solitons are roughly fitted by an expression of the form:
\begin{equation}
B=\beta\exp\left(-\frac{T^2}{w_T^2}-\frac{Y^2}{w_Y^2}\right)\sin\left[\omega (T-T_1)\right],
\end{equation}
in which the coefficients are chosen to fit the numerical data mentioned above.
For the above data, this gives the following values:
$ w_T= 0.7$, $ w_Y= 0.75$, $\omega= 3.5$,
$\beta= -18$, and $T_1=0.04$.
 \begin{figure}
\begin{center}
\includegraphics[width=7cm]{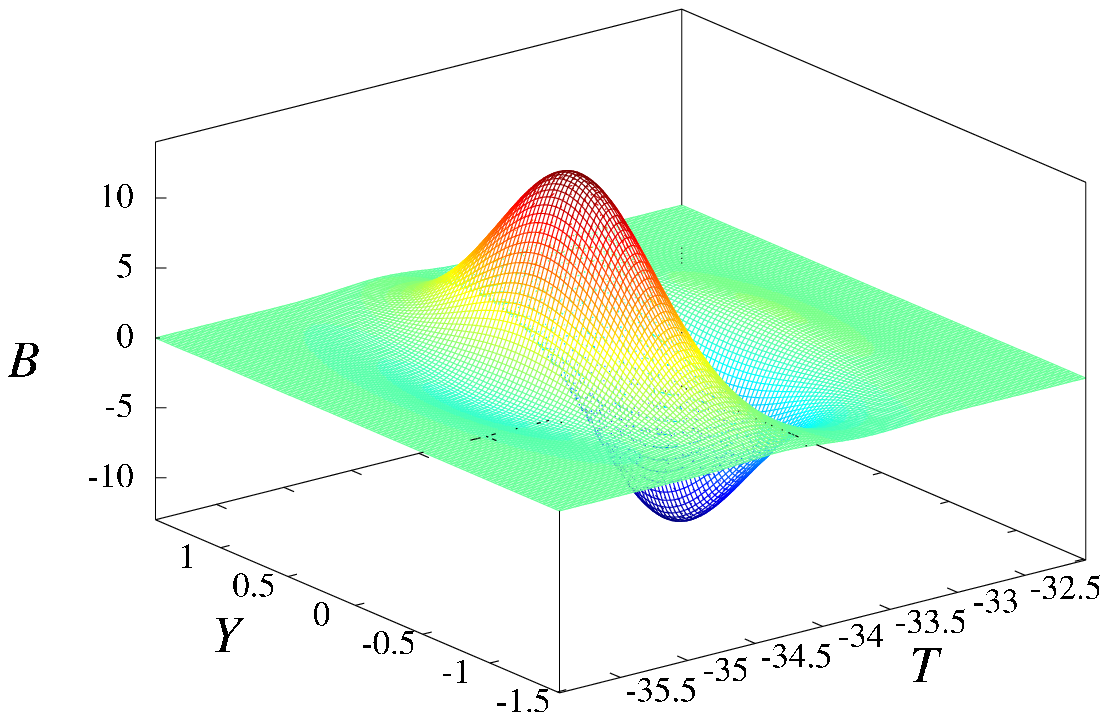}
\includegraphics[width=7cm]{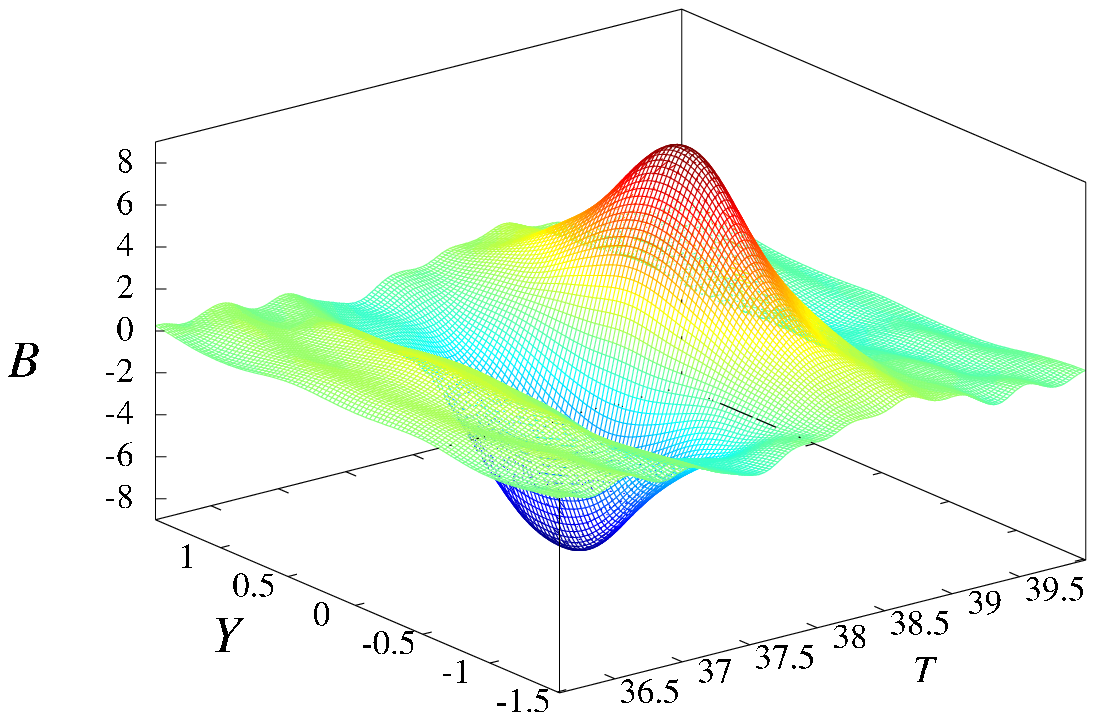}
\caption{(Color online) The evolution of a two-dimensional soliton, from an input roughly reproducing it. a) Input ($Z=0$), b) $Z=316.8$.}
\label{sg_sol}
\end{center}\end{figure}
We see that the pulse stably propagates, being neither affected by dispersion nor by diffraction; after a 
transitory stage in which the pulse radiates energy, and its amplitude decreases, stabilization is reached eventually.
The pulse energy is quite difficult to evaluate due to the oscillations of the pulse. 
A pretty good evaluation of its evolution is found by considering  the maximal value of the optical spectrum an its evolution versus $Z$. The corresponding plot is too noisy to be drawn here, but proves that stabilization occurs around $Z=100$ 
for the particular data we used.
 The nonlinear inverse velocity varies slowly from 0.65 to 1.08, and then stabilizes. The spectral width 
is also affected by the 'reshaping' and finally stabilizes, as shown in Fig. \ref{fig_spec}.
The spatial width  of pulse is not modified at all, as can be seen in Fig. \ref{fig_larg}.
Variations of the carrier-envelope phase are also observed, however they cannot be evaluated for technical reasons.

In fact, the light bullets in the 2D sG model have already been mentioned in Ref. \cite{xin00}, and even interactions have been studied in Ref. \cite{povi05}.
In \cite{xin00}, the 2D sG equation was also derived from the Maxwell-Bloch equations, but the derivation was performed from a reduced form of the Maxwell-Bloch equations, and  the physical assumptions were not so cleary given. 
We thus proved in the present work that the short wave approximation of the Maxwell-Bloch equations is indeed the 2D sG equation, and not one of the nonlinear system of equations found in the study of nonlinear electromagnetic waves in ferromagnets \cite{ferro_prl,swfer_let}.
Still in Ref. \cite{xin00}, a nonlinear Schr{\"o}dinger-type equation with higher order nonlinear terms was derived from 2D sG, and it was shown that the latter saturates the nonlinearity and may arrest collapse.
Light bullets were also simulated numerically in Ref. \cite{xin00}, but on a rather short propagation distance; it was stated in that work that the light bullets 
loose energy and will become destroyed eventually. However, according to our computations on propagation distances 20 times larger than in Ref. \cite{xin00}, in which stabilization of the energy occurs,
the energy loss seems rather to be due to the reshaping of the input pulse, and the light bullet is expected to have an infinite duration.
Later, in Ref. \cite{povi05}, the robustness of the light bullets during collisions was shown. The 2D sG equation was written in the form which allows propagation in both directions (in contrast, what we got here is the one-directional form of the 2D sG equation), and the inputs were either counterpropagating or colliding at a large angle \cite{povi05}.


\section{Conclusions}

In summary, we have introduced a generic model beyond the slowly varying envelope approximation of the nonlinear Schr{\"o}dinger-type evolution equations, for describing the propagation of (2+1)-dimensional spatiotemporal ultrashort optical solitons in two-level media. In the case of resonant nonlinearities, the short-wave approximation allow us to obtain a two-dimensional model generalizing the sine-Gordon equation. 
\cor{The use of the multiscale expansion up to the second-order in a certain small perturbation parameter allowed us to give a rigorous derivation of this equation, and more precisely to prove } that the nonlinear partial differential equation obtained in this paper is different from the previously known two-dimensional generalizations of sine-Gordon equation, see, e.g.,  \cite{ferro_prl,ferro_prb,swfer_let,swfer_perp}. 
Direct numerical simulations of the two-dimensional generalization of the sine-Gordon equation show \cor{ that stable (2+1)-dimensional ultrashort light bullets may form spontaneously from a transversely perturbed input plane wave, in the few-cycle regime.
Thus the stability of light bullets has been demonstrated, in contrast with a previous reported result \cite{xin00}. 
}  

The present study can be generalized by taking into account both resonant and non-resonant optical nonlinearities.
Moreover, the generalization to two transverse spatial dimensions, in addition to time and longitudinal coordinates, in order to study the formation of (3+1)-dimensional few-optical-cycle light bullets \cite{MMWT_review} can also be envisaged.

\end{document}